\documentclass[paper,toc,nohyper]{JHEP3}
\usepackage{cite}
\usepackage{amsbsy} 

\pdfoutput=1

\def\beq{\begin{equation}}   
\def\eeq{\end{equation}}
\def\bea{\begin{eqnarray}}  
\def\eea{\end{eqnarray}}

\def\O{y}
\def\asb{\left(\frac{\alpha_s}{2\pi}\right)}

\def\doubletilde#1{\widetilde{\vphantom{\raise 1.5pt \hbox{#1}}\smash{\kern -2pt\widetilde{#1}}}}

\def\CA{C_A}

\def\NF{N_F}

\def\mom#1{\langle #1 \rangle}

\def\d{\hbox{d}}

\def\ln{\hbox{ln}}

\title{\boldmath 
EERAD3: Event shapes and jet rates in electron-positron annihilation at order 
$\alpha_s^3$
}
\author{
A.~Gehrmann--De Ridder\\
Institute for Theoretical Physics, ETH, CH-8093 Z\"urich,
Switzerland\\ 
E-mail: \email{gehra@phys.ethz.ch}}
\author{
T.~Gehrmann\\
Physik-Institut, Universit\"at Z\"urich,
Winterthurerstrasse 190,\\ CH-8057 Z\"urich, Switzerland\\
E-mail: \email{thomas.gehrmann@uzh.ch}}
\author{E.W.N.~Glover\\
Institute for Particle Physics Phenomenology, 
        Department of Physics,\\
        University of Durham, Durham, DH1 3LE, UK\\
	E-mail: \email{e.w.n.glover@durham.ac.uk}}
\author{
G.~Heinrich\\
Max-Planck-Institut f\"ur Physik, F\"ohringer Ring 6, D-80805 M\"unchen, Germany\\
E-mail: \email{gudrun@mpp.mpg.de}}

\abstract{
The program {\tt EERAD3} computes the parton-level QCD contributions to event shapes and 
jet rates in electron-positron annihilation through to order $\alpha_s^3$. For three-jet production and 
related observables, this corresponds to next-to-next-to-leading order corrections, and allows 
for precision QCD studies. We describe the program and its usage in detail.
 }

\preprint{{ZU-TH 05/14}, {IPPP/14/15}, {MPP-2014-23}}

\begin{document}

\section{Introduction}
For hadron production at colliders, inclusive quantities 
like total cross sections 
%or sum rules 
correspond to a very simple final state definition and 
can be computed analytically to very high order in perturbation 
theory. Experimental measurements are often applying specific selection or reconstruction 
criteria to the final state, and are therefore less inclusive. Typical exclusive observables are 
jet rates or event shape distributions. 
To compute such observables 
in perturbation theory, it is necessary to implement the definition of the observable at parton-level 
in the theoretical calculation. Such a parton-level event generator includes all partonic processes 
relevant up to the required perturbative order. Each individual parton-level contribution is usually
infrared-divergent, and only the sum of all contributions produces a finite and physically well-defined 
result. To enable the implementation of the individual contributions, one typically introduces 
a  subtraction method to separate finite and divergent parts and to cancel divergences 
among different contributions. 

Perturbative calculations at next-to-leading order (NLO) in QCD, 
based on parton-level event generators, are available for a very broad spectrum of processes, 
often ranging to very high final state multiplicities. 
Substantial efforts are currently under way to extend 
calculations for low-multiplicity benchmark processes to  next-to-next-to-leading order (NNLO) in 
QCD. A pioneering calculation in this context was the NNLO corrections to event shapes 
and jet rates in electron-positron 
annihilation~\cite{GehrmannDeRidder:2007hr,GehrmannDeRidder:2008ug,weinzierl1}, 
based on the antenna subtraction method~\cite{ourant} for the handling of infrared-divergent 
kinematic contributions.  

In this paper, we describe the program {\tt EERAD3}, which we originally developed 
in the context of~\cite{GehrmannDeRidder:2007hr,GehrmannDeRidder:2008ug,GehrmannDeRidder:2007jk}. Compared
to the original version, we have made substantial improvements in efficiency and 
usability.  The program structure 
being very modular, it allows the user to either run the program for the production of 
jet rates and event shape distributions that are already implemented, or to extend the program 
towards new sets of observables. The program code can be downloaded at
{\tt http://eerad3.hepforge.org}.

\section{Event shapes and jet cross sections in perturbative QCD}

\label{sec:shapes}

Event shape observables have proven very useful
to characterise hadronic final states in electron-positron
annihilation without the need to define jets. 
These observables can be divided 
into  classes, according to the minimal number of final-state particles 
required for them to be non-vanishing: the most common variables require 
three particles (and are thus closely related to three-jet final states),
while some other variables  require 
at least four final state particles. 

Among the event shapes requiring three-particle final states, 
six ``classical" event shape variables 
can be calculated directly with the program {\tt EERAD3}.
These are the thrust $T$~\cite{farhi}, the
normalised heavy jet mass $M_H^2/s$~\cite{mh}, 
the wide and total jet
broadenings $B_W$ and $B_T$~\cite{bwbt},  
the $C$-parameter~\cite{c} and the transition from three-jet to 
two-jet final states in the Durham jet algorithm $y_{23}$~\cite{durham}.
The definitions of these variables are collected 
and described in more detail in Ref.~\cite{GehrmannDeRidder:2007hr}.

The perturbative expansion for the distribution of an 
event shape observable $\O$ up to NNLO at the centre-of-mass energy $\sqrt{s}$ 
and renormalisation scale $\mu^2 = s$, with 
$\alpha_s\equiv \alpha_s(\sqrt s)$,  is given by
\begin{eqnarray}
\frac{1}{\sigma_{{\rm had}}}\, \frac{\d\sigma}{\d \O} &=& 
\left(\frac{\alpha_s}{2\pi}\right) \frac{\d \bar A}{\d \O} +
\left(\frac{\alpha_s}{2\pi}\right)^2 \frac{\d \bar B}{\d \O}
+ \left(\frac{\alpha_s}{2\pi}\right)^3 
\frac{\d \bar C}{\d \O} + {\cal O}(\alpha_s^4)\;.
\label{eq:NNLO}
\end{eqnarray}
Here the event shape distribution  
is normalised to the total hadronic cross section $\sigma_{\rm{had}}$.
The latter can be expanded as 
  \begin{equation}
  \sigma_{\rm{had}}=\sigma_0\,
\left(1+\frac{3}{2}C_F\,\left(\frac{\alpha_s}{2\pi}\right)
+K_2\,\left(\frac{\alpha_s}{2\pi}\right)^2+{\cal O}(\alpha_s^3)\,
\right) \;,
\end{equation}
where the Born cross section for $e^+e^- \to q \bar q$ is
\begin{equation}
\sigma_0 = \frac{4 \pi \alpha}{3 s} N \, e_q^2\;,
\end{equation}
assuming massless quarks.
The constant $K_2$ is given by~\cite{chetyrkin},
\begin{equation}
  K_2=\frac{1}{4}\left[- \frac{3}{2}C_F^2
+C_FC_A\,\left(\frac{123}{2}-44\zeta_3\right)+C_FT_RN_F\,(-22+16\zeta_3)
 \right] \;,
\end{equation}
with
$\CA = N$, $C_F = (N^2-1)/(2N)$,
$T_R = {1}/{2}$,  and $N_F$ light quark flavours.

The program {\tt EERAD3} computes the perturbative coefficients $A$, $B$ and $C$, which are 
normalised to $\sigma_0$:
\begin{eqnarray}
\frac{1}{\sigma_0}\, \frac{\d\sigma}{d \O} &=& 
\left(\frac{\alpha_s}{2\pi}\right) \frac{\d  A}{\d \O} +
\left(\frac{\alpha_s}{2\pi}\right)^2 \frac{\d  B}{\d \O}
+ \left(\frac{\alpha_s}{2\pi}\right)^3 
\frac{\d  C}{\d \O} + {\cal O}(\alpha_s^4)\,.
\label{eq:NNLOsigma0}
\end{eqnarray}
$A$, $B$ and $C$ are straightforwardly related to $\bar{A}$, $\bar{B}$ 
and $\bar{C}$:
\begin{eqnarray}
&&\bar{A} = A\;,\;
\bar{B} = B - \frac{3}{2}C_F\,A\;,\;
\bar{C} = C -  \frac{3}{2}C_F\,B+ \left(\frac{9}{4}C_F^2\,-K_2\right)\,A 
\;.\label{eq:ceff}
\end{eqnarray} 
As these coefficients are computed at a renormalisation scale fixed to 
the centre-of-mass energy, they 
depend only on the value of the observable $y$.
Electroweak corrections, calculated in \cite{Denner:2009gx,Denner:2010ia}, 
are not included in the present version of {\tt EERAD3}.
Further, the pure-singlet contribution from three-gluon final states 
to three-jet observables was found to be  negligible~\cite{nigeljochum}
and is discarded. 

The QCD coupling constant evolves according to the renormalisation group 
equation, which reads to NNLO:
\begin{equation}
\label{eq:running}
\mu^2 \frac{\d \alpha_s(\mu)}{\d \mu^2} = -\alpha_s(\mu) 
\left[\beta_0 \left(\frac{\alpha_s(\mu)}{2\pi}\right) 
+ \beta_1 \left(\frac{\alpha_s(\mu)}{2\pi}\right)^2 
+ \beta_2 \left(\frac{\alpha_s(\mu)}{2\pi}\right)^3 
+ {\cal O}(\alpha_s^4) \right]\,
\end{equation}
with the following  coefficients in the $\overline{{\rm MS}}$-scheme:
\begin{eqnarray}
\beta_0 &=& \frac{11 \CA - 4 T_R \NF}{6}\;,\nonumber  \\
\beta_1 &=& \frac{17 \CA^2 - 10 C_A T_R \NF- 6C_F T_R \NF}{6}\;, \nonumber \\
\beta_2 &=&\frac{1}{432}
\big( 2857 C_A^3 + 108 C_F^2 T_R N_F -1230 C_FC_A T_R N_F
-2830 C_A^2T_RN_F \nonumber \\ &&
+ 264 C_FT_R^2 N_F^2 + 316 C_AT_R^2N_F^2\big)\;.
\end{eqnarray}

Equation~(\ref{eq:running}) 
is solved by introducing $\Lambda$ as integration constant
with $L= \log(\mu^2/\Lambda^2)$, yielding the running coupling constant:
\begin{equation}
\alpha_s(\mu) = \frac{2\pi}{\beta_0 L}\left( 1- 
\frac{\beta_1}{\beta_0^2}\, \frac{\log L}{L} + \frac{1}{\beta_0^2 L^2}\,
\left( \frac{\beta_1^2}{\beta_0^2}\left( \log^2 L - \log L - 1
\right) + \frac{\beta_2}{\beta_0}  \right) \right)\;.
\end{equation}

In terms of the running coupling $\alpha_s(\mu)$, the 
NNLO (non-singlet) expression for event shape distributions therefore becomes
\begin{eqnarray}
\frac{1}{\sigma_{{\rm had}}}\, \frac{\d\sigma}{\d \O} (s,\mu^2,\O) &=& 
\left(\frac{\alpha_s(\mu)}{2\pi}\right) \frac{\d \bar A}{\d \O} +
\left(\frac{\alpha_s(\mu)}{2\pi}\right)^2 \left( 
\frac{\d \bar B}{\d \O} + \frac{\d \bar A}{\d \O} \beta_0 
\log\frac{\mu^2}{s} \right)
\nonumber \\ &&
+ \left(\frac{\alpha_s(\mu)}{2\pi}\right)^3 
\bigg(\frac{\d \bar C}{\d \O} + 2 \frac{\d \bar B}{\d \O}
 \beta_0\log\frac{\mu^2}{s}
+ \frac{\d \bar A}{\d \O} \left( \beta_0^2\,\log^2\frac{\mu^2}{s}
+ \beta_1\, \log\frac{\mu^2}{s}   \right)\bigg)  
\nonumber \\ &&
 + {\cal O}(\alpha_s^4)\;.
\label{eq:NNLOmu} 
\end{eqnarray}

We emphasise again that the program {\tt EERAD3} computes the perturbative coefficients 
$A$, $B$ and $C$ defined in eq.~(\ref{eq:NNLOsigma0}), where the renormalisation scale 
has been fixed to the centre-of-mass energy. 
However, for convenience of the user, we include an auxiliary program 
{\tt eerad3\_dist.f} in the program package, which produces results according to 
eq.~(\ref{eq:NNLOmu}) and performs scale variations. More details about the usage of 
{\tt eerad3\_dist.f}  are given in Section~\ref{sec:eerad3dist}.

The program {\tt EERAD3} has been used to compute event-shape
distributions~\cite{GehrmannDeRidder:2007hr} and their moments~\cite{ourmom} as well as jet 
rates~\cite{GehrmannDeRidder:2008ug}. An independent 
validation of these results (also using the antenna 
subtraction method), resolving problems with large-angle 
soft terms~\cite{weinzierl}, was made in~\cite{weinzierl1,weinzierl2,Weinzierl:2009yz,Weinzierl:2010cw}. 
These results enabled a substantial number of 
precision QCD studies based on the reanalysis of LEP 
data~\cite{jadeas,Schieck:2012mp,ouras,jetsas,opalas,momas,asthrust1,asthrust2,asthrust3,Pahl:2009aa,Abbate:2012jh}.
In the dijet-limit, resummation techniques predict the logarithmically 
enhanced terms in the event shape distributions. These were derived 
independently~\cite{asthrust1,mhresum,monni,btbwresum} and 
served as validation of the  {\tt EERAD3} results.

\section{Structure of the program {\tt EERAD3}}
\label{sec:structure}

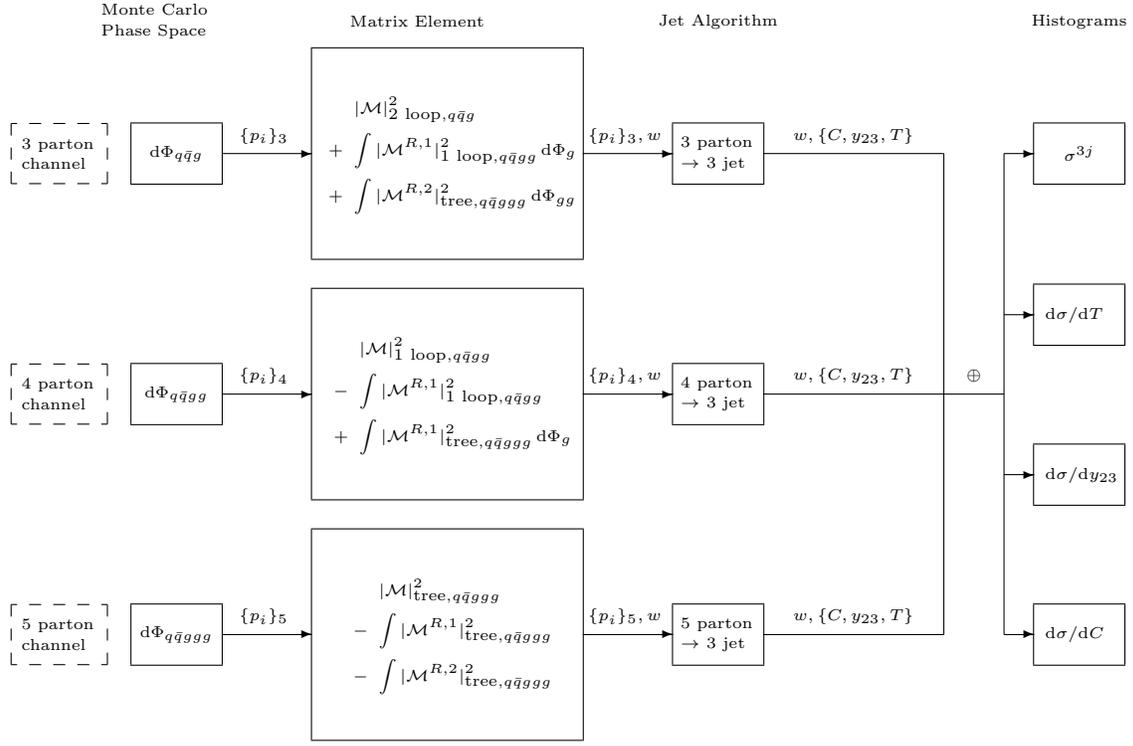
\begin{figure}
\hspace{-0.7cm}
\unitlength 0.8cm
{\tiny 
\begin{picture}(28,15)
\put(1,4){\dashbox{0.2}(1.6,1){\hspace{0.3cm}\parbox{1.3cm}{5 parton\\channel}}}
\put(1,8){\dashbox{0.2}(1.6,1){\hspace{0.3cm}\parbox{1.3cm}{4 parton\\channel}}}
\put(1,12){\dashbox{0.2}(1.6,1){\hspace{0.3cm}\parbox{1.3cm}{3 parton\\channel}}}
\put(3,4){\framebox(1.5,1){$\d \Phi_{q\bar q ggg}$}}
\put(3,8){\framebox(1.5,1){$\d \Phi_{q\bar q gg}$}}
\put(3,12){\framebox(1.5,1){$\d \Phi_{q\bar q g}$}}
\put(2.75,14.2){\makebox(2,1){\parbox{2cm}{Monte Carlo\\Phase Space}}}
\put(6,2.75){\framebox(4.5,3.5){\parbox{4cm}{\begin{eqnarray*}
&&|{\cal M}|^2_{\mbox{tree}, q \bar q ggg}\\
&-& \int |{\cal M}^{R,1}|^2_{\mbox{tree}, q \bar q ggg} \\
&-& \int |{\cal M}^{R,2}|^2_{\mbox{tree}, q \bar q ggg} 
\end{eqnarray*}}}}
\put(6,10.75){\framebox(4.5,3.5){\parbox{4cm}{\begin{eqnarray*}
&&|{\cal M}|^2_{\mbox{2 loop}, q \bar q g}\\
&+& \int |{\cal M}^{R,1}|^2_{\mbox{1 loop}, q \bar q gg} \, \d \Phi_g\\
&+& \int |{\cal M}^{R,2}|^2_{\mbox{tree}, q \bar q ggg} \, \d \Phi_{gg}
\end{eqnarray*}}}}
\put(6,6.75){\framebox(4.5,3.5){\parbox{4cm}{\begin{eqnarray*}
&&|{\cal M}|^2_{\mbox{1 loop}, q \bar q gg}\\
&-& \int |{\cal M}^{R,1}|^2_{\mbox{1 loop}, q \bar q gg} \\
&+& \int |{\cal M}^{R,1}|^2_{\mbox{tree}, q \bar q ggg} \, \d \Phi_g
\end{eqnarray*}}}} 
\put(4.5,4.5){\vector(1,0){1.5}}
\put(4.7,4.5){\makebox(1,0.6){$\{p_i\}_5$}}
\put(4.5,8.5){\vector(1,0){1.5}}
\put(4.7,8.5){\makebox(1,0.6){$\{p_i\}_4$}}
\put(4.5,12.5){\vector(1,0){1.5}}
\put(4.7,12.5){\makebox(1,0.6){$\{p_i\}_3$}}
\put(7,14.2){\makebox(1.5,1){Matrix Element}}
\put(10.5,4.5){\vector(1,0){1.5}}
\put(10.7,4.5){\makebox(1,0.6){$\{p_i\}_5,w$}}
\put(10.5,8.5){\vector(1,0){1.5}}
\put(10.7,8.5){\makebox(1,0.6){$\{p_i\}_4,w$}}
\put(10.5,12.5){\vector(1,0){1.5}}
\put(10.7,12.5){\makebox(1,0.6){$\{p_i\}_3,w$}}
\put(12,14.2){\makebox(1.5,1){Jet Algorithm}}
\put(12,4){\framebox(1.5,1){\hspace{0.3cm}\parbox{1.3cm}{5 parton\\$\rightarrow$ 3 jet}}}
\put(12,8){\framebox(1.5,1){\hspace{0.3cm}\parbox{1.3cm}{4 parton\\$\rightarrow$ 3 jet}}}
\put(12,12){\framebox(1.5,1){\hspace{0.3cm}\parbox{1.3cm}{3 parton\\$\rightarrow$ 3 jet}}}
\put(13.5,4.5){\line(1,0){3}}
\put(14,4.5){\makebox(2,0.6){$w,\{C,y_{23},T\}$}}
\put(13.5,8.5){\line(1,0){3}}
\put(14,8.5){\makebox(2,0.6){$w,\{C,y_{23},T\}$}}
\put(13.5,12.5){\line(1,0){3}}
\put(14,12.5){\makebox(2,0.6){$w,\{C,y_{23},T\}$}}
\put(16.5,12.5){\line(0,-1){8}}
\put(17.5,12.5){\line(0,-1){8}}
\put(17.5,12.5){\vector(1,0){0.5}}
\put(17.5,7.16){\vector(1,0){0.5}}
\put(17.5,9.82){\vector(1,0){0.5}}
\put(17.5,4.5){\vector(1,0){0.5}}
\put(16.5,8.5){\line(1,0){1}}
\put(16.5,8.5){\makebox(1,0.6){$\oplus$}}
\put(18,14.2){\makebox(1.5,1){Histograms}}
\put(18,12){\framebox(1.5,1){\hspace{0.3cm}\parbox{0.7cm}{ $ \sigma^{3j} $}}}
\put(18,9.32){\framebox(1.5,1){\hspace{0.3cm}\parbox{1.2cm}{ $\d \sigma/\d T $}}}
\put(18,6.66){\framebox(1.5,1){\hspace{0.3cm}\parbox{1.2cm}{ $\d \sigma/\d y_{23} $}}}
\put(18,4){\framebox(1.5,1){\hspace{0.3cm}\parbox{1.2cm}{ $\d \sigma/\d C $}}}
\end{picture}}\vspace{-2cm}
\caption{Schematic structure of {\tt EERAD3}. The event shape observables $C,y_{23},T$ are for
illustration, other observables can be calculated as well.}
\label{fig:mc}
\end{figure}

The program {\tt EERAD3} computes the three-parton, four-parton and five-parton 
contributions to hadronic final states in $e^+e^-$ annihilation, and combines them with 
subtraction terms appropriate to NNLO in QCD. Any infrared-safe quantity derived 
from three-particle final states can thus be computed with this program to ${\cal O}(\alpha_s^3)$. 
The structure of the program is depicted in Figure~\ref{fig:mc}.

The source code of {\tt EERAD3} consists of the following files:
\begin{tabbing}
{\tt eerad3.f}: \hspace{2cm}\= the main program\\
{\tt 3jme.f}: \>  two-loop three-parton matrix elements, from~\cite{3jme} \\
{\tt aversub0.f}: \> subroutines for subtraction at NLO\\
{\tt aversub1.f}: \> subroutines for subtraction of double real radiation at NNLO\\
{\tt aversub2.f}: \> subroutines for subtraction of one-loop real radiation at NNLO\\
{\tt brem.f}: \> real radiation matrix elements\\
{\tt ecuts.f}: \> jet algorithms and event shape definitions\\
{\tt eerad3lib.f}: \> library with special functions and auxiliary routines\\
{\tt histo.f}: \> histogram handling routines\\
{\tt hplog.f}: \> one-dimensional harmonic polylogarithms, from~\cite{Gehrmann:2001pz}\\
{\tt phaseee.f}: \> phase space routines\\
{\tt sig.f}: \> parton-level cross sections and subtraction terms\\
{\tt tdhpl.f}: \> two-dimensional harmonic polylogarithms, from~\cite{Gehrmann:2001jv}\\
{\tt virt.f}: \> one-loop four-parton matrix elements, from~\cite{fourjet}
\end{tabbing}
Their content and function is outlined in the following.
Two auxiliary programs are
{\tt eerad3\_combine.f} and {\tt eerad3\_dist.f}.

\subsection{Main program}
The main program {\tt eerad3.f} is 
steering the input/output and the Monte Carlo integration. 
The subroutine {\tt readinit} reads the input file {\tt eerad3.input} and initialises the settings 
for the colour factors and observables to be calculated, as well as for the 
Monte Carlo integration. 
The calculation of  the different parts contributing to the NNLO cross section 
is performed through a call to the subroutine {\tt cross}. In this subroutine, 
the Monte Carlo integration is done with {\tt vegas}~\cite{vegas}, proceeding in two steps: 
first a grid is constructed, then the actual integration is performed based on this grid.
The grid step can also be skipped once a grid is produced and saved to a file by setting {\tt
iwarm=0} in {\tt eerad3.input}. Likewise, {\tt iprod=0} skips the integration and production of the histograms. 
The {\tt vegas} integration routines, adapted for the program {\tt EERAD3}, are defined in the file 
{\tt eerad3lib.f}.

\subsection{Cross section}
The cross section consists of three basic parts 
which correspond to different particle multiplicities in the final state 
and therefore are integrated separately. 
\begin{itemize}
\item {\tt sig3} corresponds to the leading order kinematics with three identified particles 
in the final state. This part contains the Born cross section as well as the 
one-loop and two-loop virtual
corrections (from {\tt 3jme.f}, using {\tt hplog.f} and {\tt tdhpl.f})
and the finite remainders of the integrated subtraction terms 
cancelling the poles
of the two-loop virtual corrections.
\item {\tt sig4} corresponds to four particle kinematics at tree-level or one loop 
(from {\tt virt.f}),  
where one of the final state particles can be theoretically unresolved (soft or collinear), 
and therefore requires the inclusion of infrared subtraction terms. 
\item {\tt sig5} corresponds to five particle kinematics (from {\tt brem.f}),  
where two of the final state particles can be theoretically unresolved, 
requiring the inclusion of infrared subtraction terms for unresolved double real
radiation.
\end{itemize} 
This basic structure is reflected by the functions {\tt sig3a, sig4a, sig5a} defined in 
the file {\tt sig.f}. We documented the expressions implemented for matrix elements and 
subtraction terms in these functions in~\cite{GehrmannDeRidder:2007jk}. 

The subtraction terms are contained in the files {\tt aversub*.f}, where {\tt aversub0.f}
contains the antenna functions and momentum mappings for the singly unresolved parts, while 
{\tt aversub1.f} contains the functions and momentum mappings for the doubly unresolved 
subtraction terms and  {\tt aversub2.f} for the single unresolved subtraction terms at one loop. 
The one-loop and two-loop matrix elements~\cite{3jme} are expressed in terms of 
one-dimensional~\cite{hpl} and two-dimensional~\cite{tdhpl} harmonic polylogarithms. These 
functions are evaluated in {\tt hplog.f}~\cite{Gehrmann:2001pz} and
 {\tt tdhpl.f}~\cite{Gehrmann:2001jv}.

\subsection{Phase space integration}
Based on the {\tt vegas} integration variables, 
the routines in {\tt phaseee.f} generate phase space points for three, four and five partons. 
The four-parton and five-parton phase spaces are decomposed into wedges (6 wedges 
at the four-parton level and 45 wedges at the five-parton level). Each wedge contains only 
certain classes of unresolved limits, such that a phase space parametrisation appropriate 
to these  limits can be used. At the four-parton level, permutations of 
a  single parametrisation are sufficient, while 
two parametrisations (double single collinear (two pairs of collinear partons)
and triple collinear (three collinear partons)) are used at 
the five-parton level. An angular rotation of unresolved pairs of momenta is performed 
inside each wedge, such that angular 
correlations~\cite{weinzierlang,cs,catanigrazzini}
 in the collinear splitting functions can be 
cancelled out by averaging. The four-parton phase space contains at most single collinear limits, 
such that two angular-correlated phase space points are sufficient. Four angular-correlated phase 
space points are generated for the five-parton phase space to account for angular terms in 
double single collinear and triple collinear limits. 

\subsection{Event shapes and jet cross sections}
Based on the four-momenta provided by the phase space generator (or the phase space mappings), 
parton-level values for event shapes and jet transition parameters are computed in {\tt ecuts.f}.
To add further shape variables or different jet algorithms, the user can extend these routines.  
The parton momenta are contained in {\tt ppar(i,j)}, where {\tt i=(1:4)} enumerates the momentum 
components (with energy in the fourth entry) and {\tt j=(1:5)} enumerates the momenta. For 
each event shape variable or jet transition rate, infrared-safe implementations for 
{\tt npar=3,4,5}, the number of partons in the final state, 
must be provided. 

In {\tt ecuts.f}, events are accepted if they pass the cuts for at least one of the distributions that 
are computed. Different distributions can be computed simultaneously, such that the result of the 
vegas integration will usually not have a physical meaning (except if the 
program is run for a single observable, potentially selecting a specific moment weight). Instead, 
the physical distributions are contained in the histograms generated by the program. 

\subsection{Calculation of moments of event shape observables}
\label{sec:moments}

The $n$-th moment of an event shape observable $y$ is
defined as 
\begin{equation}
\mom{y^n}=\frac{1}{\sigma_{\rm{had}}}\,\int_0^{y_{\rm{max}}} y^n 
 \frac{\d\sigma}{\d y} \d y \;,
\end{equation}
where $y_{\mathrm{max}}$ is the kinematically allowed upper limit of the
observable.  
As the calculation of moments involves an integration over the full phase
space,  they offer a way of studying an observable which is complementary to the use
of distributions; in particular, they are useful to 
investigate non-perturbative effects. 
Recent studies of event shape moments 
can be found 
in~\cite{ourmom,Weinzierl:2009yz,momas,Abbate:2012jh,Pahl:2008uc,Pahl:2009aa}.

The  perturbative QCD expansion of $\mom{y^n}$ is given by~\cite{ourmom}
\begin{equation}
  \mom{y^n}(s,\mu^2 = s) = \asb \bar{{\cal A}}_{y,n}  + 
\asb^2 \bar{{\cal B}}_{y,n} + \asb^3 \bar{{\cal C}}_{y,n} +
  {\cal O}(\alpha_s)^4\;.
\label{eq:pertexp}
\end{equation}
Just as in eq.~(\ref{eq:NNLOmu}), the 
NNLO expression for an event shape moment measured at centre-of-mass 
energy squared $s$ evaluated at renormalisation scale $\mu$ becomes,  
\begin{eqnarray}
\mom{y^n} (s,\mu^2) &=& 
\left(\frac{\alpha_s(\mu)}{2\pi}\right) \bar{{\cal A}}_{y,n} +
\left(\frac{\alpha_s(\mu)}{2\pi}\right)^2 \left( 
\bar{{\cal B}}_{y,n}+ \bar{{\cal A}}_{y,n} \beta_0 
\log\frac{\mu^2}{s} \right)
\nonumber \\ &&
+ \left(\frac{\alpha_s(\mu)}{2\pi}\right)^3 
\bigg(\bar{{\cal C}}_{y,n}+ 2 \bar{{\cal B}}_{y,n}
 \beta_0\log\frac{\mu^2}{s}
+ \bar{{\cal A}}_{y,n} \left( \beta_0^2\,\log^2\frac{\mu^2}{s}
+ \beta_1\, \log\frac{\mu^2}{s}   \right)\bigg)  
\nonumber \\ &&
 + {\cal O}(\alpha_s^4)\;.
\label{eq:momNNLOmu} 
\end{eqnarray}
To calculate a particular moment for a particular event shape variable ({\tt iaver=1..7}), 
{\tt EERAD3} must be run with 
{\tt imom} set to the desired moment number. The result is then obtained as the
integration output. 
Note that the default setting {\tt imom=1} should be used to calculate  distributions
and jet cross sections.

\subsection{Booking of results into histograms}
\label{sec:histo}

The histograms are defined in {\tt histo.f}, using a generic, observable-independent histogram manager.
The latter is defined in the subroutine {\tt ghiman} in {\tt eerad3lib.f}. The communication with the 
histogram manager is made through a subroutine {\tt bino}, which can be extended by the user. 
For each histogram, four calls to the histogram manager are specified in {\tt bino}; {\tt histoi} initialises 
the histogram, {\tt histoa} adds an event to the histogram, {\tt histoe} computes 
the histogram errors and {\tt histow} writes out the final histograms.
Each histogram is identified by its identifier number {\tt histoID}, its minimum and maximum bin 
values {\tt bmin/bmax}  and its number of bins {\tt nbins}. 
To change histogram bins and boundaries, or to introduce histograms for new variables, the user can modify the appropriate calls 
in {\tt bino}. 

The {\tt EERAD3} program produces histograms with linear or logarithmic binning. Since 
the phase space requirements and variable ranges 
differ substantially for the different binnings, 
they can not be produced in a single run. For $0\leq${\tt iaver}$\leq 5$, only the 
linearly binned histograms are computed, while $6\leq${\tt iaver}$\leq 8$ produces only histograms with 
logarithmic binning. 

For each event shape variable, three types of histograms can be produced:
in the case of the thrust distribution in $\tau=1-T$, these are for example  
\begin{enumerate}
\item $\langle\tau\rangle$ distribution $\frac{\tau}{\sigma}\,\frac{d\sigma}{d\tau}$ for four different binnings, 
{\tt nbin=200,100,50,25}, with {\tt bmin=0, bmax=0.5}. Produced for {\tt iaver=0} or {\tt iaver=4}.
\item  $\tau$ distribution $\frac{1}{\sigma}\,\frac{d\sigma}{d\tau}$ for four different binnings, 
{\tt nbin=200,100,50,25}, with {\tt bmin=0, bmax=0.5}. Produced for {\tt iaver=0} or {\tt iaver=4}.
\item $-\ln(\tau)$ distribution $\frac{1}{\sigma}\,\frac{d\sigma}{d(-\ln(\tau))}$ for three different binnings, 
{\tt nbin=100,50,25}, with {\tt bmin=0, bmax=10}. Produced for {\tt iaver=8}.
\end{enumerate}
The $C$-parameter has different default linear binnings, {\tt nbin=400,200,100,50}. 

The jet rates and transition parameters are always 
booked into logarithmic histograms, which means that
the produced histograms will be
$R_3(-\ln(y_{{\rm cut,D}})) = \frac{1}{\sigma}\sigma_3(-\ln(y_{{\rm cut,D}}))$ for the 
$y_{{\rm cut}}$ dependence
of the three-jet rate with the Durham $k_t$ jet algorithm~\cite{durham}, 
$\frac{1}{\sigma}\,\frac{d\sigma}{d(-\ln(y_{23,D}))}$, for the 2-to-3-jet transition 
parameter with the Durham $k_t$ jet algorithm, analogous for the 3-to-4-jet
and 4-to-5-jet transition parameters $y_{34,D}$ and $y_{45,D}$. 
These histograms are produced using {\tt iaver=6}, 
while {\tt iaver=7} produces the same distributions with the Jade 
jet algorithm~\cite{Bethke:1991wk}. 
For {\tt iaver=8} the output consists of
all jet distributions based on the Durham $k_t$ jet algorithm and all logarithmic event shape distributions. 

To add additional histograms into {\tt bino}, the user should initialise the histograms with a call to 
{\tt histoi(histoID,bmin,bmax,nbins)}, and fill the histogram with the corresponding variable 
value {\tt val} 
and Monte Carlo weight {\tt wgt} with a call to {\tt histoa(histoID,val,wgt)}.
The subroutine {\tt histoe(histoID)} calculates the statistical errors after each
{\tt vegas} iteration, and
routine {\tt histow(histoID)} writes the histogram to the screen,
output to a file with unit number {\tt lun} is done with 
 {\tt histowf(histoID,lun)}.
All these subroutines are defined in {\tt eerad3lib.f}.

If the user would like to modify or make additions to the default observable definitions 
or jet algorithms, the file {\tt ecuts.f} should be edited.
The jets are defined in the subroutine {\tt getjet}. The Durham $k_t$ and Jade 
algorithms are implemented with different options for the recombination schemes.
The default for the Durham $k_t$  algorithm is the so-called ``E-scheme"~\cite{Bethke:1991wk}, where 
the sum of two momentum vectors is defined by the usual vector sum,
treating the energy and spatial components on the same footing.
The default for the Jade algorithm is the ``E0-scheme"~\cite{Bethke:1991wk}, where 
only the energy components are simply summed, while the sum of the spatial components
is weighted by a factor. The latter is given by the sum of the energy components 
divided by the modulus of the sum of the spatial components.
The different recombination schemes are defined in the subroutine {\tt jetco}, 
which is called by {\tt getjet}. 
All these subroutines, as well as the definitions of the event shape observables, 
are defined in the file {\tt ecuts.f}. 

The cuts which are read from the input card  are set in the 
subroutine {\tt readinit} in the main program {\tt eerad3.f} and 
shared with the subroutine {\tt ecuts} in {\tt ecuts.f}. 

The file {\tt eerad3\_combine.f} allows to combine results from several different runs into a single 
set of histograms as described in section \ref{sec:combine}.
While {\tt EERAD3} computes only the perturbative coefficients $A,B,C$, the program 
{\tt eerad3\_dist.f} can be used to construct cross sections and distributions according to 
(\ref{eq:NNLOmu}), and to perform scale variations. 
Detailed instructions for the usage of {\tt eerad3\_combine.f} and 
{\tt eerad3\_dist.f}  are given in Sections~\ref{sec:combine} and \ref{sec:eerad3dist}, respectively.

\section{Usage of {\tt EERAD3}}

\subsection{Main program {\tt eerad3}}
\label{sec:eerad3}

The main program {\tt EERAD3} computes individual colour structures of the perturbative coefficients
of event shapes and jet rates. It is controlled through an input card. The program can be compiled by simply 
executing the command {\tt make}, which will produce an executable called {\tt eerad3}. 
The calculation can be started with command line options 
\begin{verbatim}
$ eerad3 -i filename.input -n XX
\end{verbatim}
where {\tt filename.input} is the name of the input card file and {\tt XX} is an integer 
between 0 and 99 selecting the random seed for runs with independent statistics.
Default values {\tt filename.input=eerad3.input} and {\tt XX=0} are inserted if no command line options 
are given. 

\subsection*{Structure of the input card}

A typical example for an input card looks as follows:
\begin{verbatim}
1d-5      ! y0
0         ! iaver
0.0025    ! cutvar
1         ! imom
1         ! iang
-2        ! nloop
1         ! icol
Z         ! ichar
1 1       ! iwarm iprod
5 5       ! itmax1 itmax2
5000 30000 200000    ! nshot3 nshot4 nshot5
\end{verbatim}
The individual entries are:
\begin{tabbing}
{\tt y0}:\hspace{2cm} \= technical cut-off for the phase space integration (dimensionless), \\ 
\> should be between $10^{-5}$ and $10^{-8}$. \\
{\tt iaver}: \> selects the observables to be computed:\\
\> 0: all event shape distributions $(B_W, C, M_H^2/s, (1-T), B_T)$, \\ 
\> 1: wide jet broadening $B_W$.\\
\> 2: $C$-parameter $C$.\\
\> 3: heavy jet mass $M_H^2/s$.\\
\> 4: thrust $1-T$.\\
\> 5: total jet broadening: $B_T$.\\
\> 6: jet rates and transition parameters in the Durham $k_t$ algorithm:\\   
\> \phantom{6: }$R_3$, $R_4$, $R_5$, $y_{23}$, $y_{34}$, $y_{45}$.\\
\> 7:  jet rates and transition parameters in the Jade algorithm:\\ 
\> \phantom{7: }$R_3$, $R_4$, $R_5$, $y_{23}$, $y_{34}$, $y_{45}$.\\
\> 8: all jet distributions in the Durham $k_t$ algorithm and all\\ 
\> \phantom{8: }logarithmic event shape distributions.\\
{\tt cutvar}: \> lower cut-off on distributions, should be at least one order of magnitude
\\ \> larger than {\tt y0}.\\
{\tt imom}: \> moment number applied as weight: if computing distributions, \\
\> should be set to {\tt imom=1} \\
{\tt iang}: \> angular optimisation of phase space: on ({\tt iang} = 1) or off  ({\tt iang} = 2).\\
{\tt nloop}: \> perturbative order: LO ({\tt nloop}=0), NLO ({\tt nloop}=-1), NNLO ({\tt nloop}=-2).\\
{\tt icol}: \> colour factor:\\
\> 0: sum all colour factors,\\
\> 1: NLO $N$,\\
\> 2: NLO $1/N$,\\
\> 3: NLO $N_F$,\\
\> 1: NNLO $N^2$,\\
\> 2: NNLO $N^0$,\\
\> 3: NNLO $1/N^2$,\\
\> 4: NNLO $N_FN$,\\
\> 5: NNLO $N_F/N$,\\
\> 6: NNLO $N_F^2$.\\
{\tt ichar}: \> one character to identify output files.\\
{\tt iwarm}: \> produce a warm up integration grid ({\tt iwarm}=1) or read the grid \\
\> from files ({\tt iwarm}=0).\\
{\tt iprod}: \> produce histograms: yes ({\tt iprod}=1) or no ({\tt iprod} = 0).\\
{\tt itmax1,2}: \> number of iterations for warm up and production runs.\\
{\tt nshot3,4,5}: \> number of vegas points for three-parton, four-parton, 
five-parton channels.
\end{tabbing}

\subsection*{Calculation of moments of  event shapes}
%\label{sec:usage_moments}

As mentioned already in Section \ref{sec:moments}, the $n$-th moment of an event shape observable 
can be calculated by selecting {\tt imom=}\,$n$ in the input card.
Please note that for the calculation of the moments the options  {\tt iaver=0} and {\tt iaver=8}
can not be used. These options  calculate a whole set of observables, using an 
unphysical 
integrand for the {\tt vegas} integration, which is appropriately 
re-weighted only for the booking of events into different histograms.

\subsection*{Structure and naming conventions of the output files}

The column structure of the output histogram files is as follows:
\begin{verbatim}
variable       observable     error
\end{verbatim}
The filenames are composed as
\begin{verbatim}
         E[aa].y[bbb].i[c][d].[e][f][g]
\end{verbatim}
with
\begin{tabbing}
{\tt [aa]}:\hspace{1cm} \= two-digit identifier for random seed (input from command line)\\
{\tt [bbb]}: \> value of {\tt y0} in format {\tt ndi}= {\tt n}$\cdot 10^{{\tt i}}$.\\
{\tt [c]}: \> one-character filename identifier {\tt ichar}.\\
{\tt [d]}: \> one-digit colour factor identifier {\tt icol}.\\
{\tt [e]}: \> one-character identifier of observable selected by {\tt iaver}, \\
           \> where the conventions are: \\
	   \> W: $B_W$\\ 
	   \> C: $C$ \\
	   \> M: $M_H^2/s$ \\
	   \> T: $\tau$ \\
	   \> B: $B_T$ \\
	   \> Y: jet transition parameters $y_{ij}$ \\
	   \> S: jet rates $y_{n}$ \\
{\tt [f]}: \> one-character identifier of distribution type,\\
\> for event shapes:\\
\> {\tt 1}: distribution $1/\sigma_0\, y ({\rm d} \sigma/{\rm d} y)$\\
\> {\tt 2}: distribution $1/\sigma_0\,  ({\rm d} \sigma/{\rm d} y)$\\
\> {\tt L}: distribution $1/\sigma_0\,  ({\rm d} \sigma/{\rm d} L)$ with $L=\ln y$\\
\> for jet rates and transition parameters (logarithmic binning):\\
\> {\tt 3}: three-jet rate and $y_{23}$ transition parameter\\
\> {\tt 4}: four-jet rate and $y_{34}$ transition parameter\\
\> {\tt 5}: five-jet rate and $y_{45}$ transition parameter\\
{\tt [g]}: \> one-character identifier of bin size (resolution), runs from {\tt a} to {\tt d}, \\
\> corresponding to the binnings described in Section \ref{sec:histo}.\\
\end{tabbing}

\subsection{{\tt eerad3\_combine}}
\label{sec:combine}
To obtain statistically independent samples, {\tt eerad3} can be run with different random seeds. 
The combination of different runs is performed by the program {\tt eerad3\_combine}, which reads an 
input card
\begin{verbatim}
$ eerad3_combine -i filename.input
\end{verbatim}
The default input filename is {\tt eerad3\_combine.input}. The content of this file is as follows
\begin{verbatim}
0       ! iaver
y1d5    ! frooty
iZ3     ! frooti
tx      ! filetag
1 20    ! minfile maxfile
3       ! nvoid
2       ! ivoid(1)
6       ! ivoid(2)
14      ! ivoid(3)
\end{verbatim}
Where:
\begin{tabbing}
{\tt iaver}:\hspace{2cm} \= selects observables to be computed, as above.\\
{\tt frooty}: \> four-character filename extension to indicate the value of {\tt y0},\\
\> i.e. {\tt y[bbb]} described above in the filename composition.\\
{\tt frooti}: \> three-character filename extension to indicate identifier and colour factor.\\
\> i.e. {\tt i[c][d]} described above in the filename composition.\\
{\tt filetag}: \> two-character identifier for combination files.\\
{\tt minfile}, {\tt maxfile}: \> random seed identifiers to fix range of files to be combined. \\
{\tt nvoid}: \> number of (void or unused) runs to be excluded from the combination.\\
{\tt ivoid(1:nvoid)} \> random seed identifiers of files to be excluded.
\end{tabbing}
The syntax of the combination filenames is as for the individual runs, with the random 
seed identifier number replaced by {\tt filetag}.

\subsection{{\tt eerad3\_dist}}
\label{sec:eerad3dist}

The main program {\tt eerad3} computes the perturbative coefficients of event shape 
distributions and jet cross sections, as defined in (\ref{eq:NNLOsigma0}). Cross sections 
and distributions according to eq.~(\ref{eq:NNLOmu})
are obtained from these coefficients using {\tt eerad3\_dist}, which read an 
input card
\begin{verbatim}
$ eerad3_dist -i filename.input
\end{verbatim}
The default input filename is {\tt eerad3\_dist.input}. The content of this file is as follows
\begin{verbatim}
8                ! iaver
E00.y1d8.iL0     !   LO
Etx.y1d8.iN1     !  NLO icol=1
Etx.y1d8.iN2     !  NLO icol=2
Etx.y1d8.iN3     !  NLO icol=3
Etx.y1d6.iZ1     ! NNLO icol=1
Etx.y1d6.iZ2     ! NNLO icol=2
Etx.y1d6.iZ3     ! NNLO icol=3
Etx.y1d6.iZ4     ! NNLO icol=4
Etx.y1d6.iZ5     ! NNLO icol=5
Etx.y1d6.iZ6     ! NNLO icol=6
EEt.091.1189     ! outfile
0.1189d0         ! alphas(MZ)
91.1889d0        ! MZ
91.1889d0        ! roots
1.d0             ! xmu
\end{verbatim}
Where:
\begin{tabbing}
{\tt iaver}:\hspace{2cm} \= selects observables to be computed, as above.\\
{\tt LO,...}: \> names (without extensions) of the files containing the histograms;\\
\> fixed to 12 characters by construction.\\
{\tt outfile}: \> name (without extensions) for output files.\\
{\tt alphas(MZ)}: \> value of the strong coupling constant $\alpha_s(M_Z)$.\\
{\tt MZ}: \> mass of $Z$-boson.\\
{\tt roots}: \> $e^+e^-$ centre-of-mass energy $\sqrt{s}$.\\
{\tt xmu}: \> default value of renormalisation scale: $\mu = x_{\mu} \sqrt{s}$. 
\end{tabbing}

The following files are produced:
\begin{tabbing}
{\tt outfile.xNNLO.[e][f][g]}:\hspace{2cm} \= distributions up to NNLO.\\
{\tt outfile.mudep.[e][f][g]}:\> distributions with range of scale variation.\\
{\tt outfile.muran.[e][f][g]}:\> percentages of scale variation.\\
{\tt outfile.histA.[e][f][g]}:\> perturbative coefficient $\bar{A}$.\\
{\tt outfile.histB.[e][f][g]}:\> perturbative coefficient $\bar{B}$.\\
{\tt outfile.histC.[e][f][g]}:\> perturbative coefficient $\bar{C}$.
\end{tabbing}
The column structure of the {\tt xNNLO} files is as follows:
\begin{verbatim}
variable    sigmaLO       sigmaNLO        sigmaNNLO      errNNLO
\end{verbatim}
In here {\tt errNNLO} is the numerical integration error at NNLO. 

The scale variation uncertainty is determined by finding the minimum and maximum of the cross 
section over a range $\mu/2$ to $2\mu$ around the default value $\mu$. To resolve extrema 
inside this 
interval, 20 logarithmically spaced points are computed. The results are contained in
the {\tt mudep} files, which have the following structure:
\begin{verbatim}
variable  sigLO   errLO   sigNLO   errNLO   sigNNLO   errNNLO
\end{verbatim}
The errors here are the errors from scale variations. 

To quantify the precision of the predictions, scale uncertainties in per cent are provided in the 
 {\tt muran} files, with the conventions:
\begin{verbatim}
variable   %(LO)       %(NLO)      %(NNLO)
\end{verbatim}

The scale variation errors in the {\tt mudep} and  {\tt muran} files do not take proper account of 
the integration errors on the NNLO coefficients. For a full consideration of error propagation,
it is therefore recommended to produce {\tt xNNLO} files with different values of {\tt xmu}. 

The coefficients $\bar{A}$, $\bar{B}$ and $\bar{C}$ are defined in (\ref{eq:ceff}), the structure of 
the {\tt histoA}, {\tt histoB} and {\tt histoC} files is as described in Section~\ref{sec:eerad3} above. 

\section{Summary}

The program {\tt EERAD3} computes jet cross sections and event shapes in 
electron-positron annihilation to order $\alpha_s^3$, corresponding to NNLO accuracy 
in perturbative QCD. The standard set of event shape observables and jet rates 
studied at LEP is implemented in the program. 
We have documented the structure and 
usage of {\tt EERAD3} in detail, and explained how the user can extend the program to 
compute other shape variables or use alternative jet algorithms. The program is available at 
{\tt http://eerad3.hepforge.org}.

\section*{Acknowledgements}

GH would like to thank the Physics Institute at the  University of Zurich
for hospitality while parts of this project were carried out.
This research was supported in part by the UK Science and Technology Facilities
Council,  in part by the Swiss National Science Foundation (SNF) under contracts
PP00P2-139192 and 200020-149517 and in part by the European
 Commission through the ``LHCPhenoNet"
Initial Training Network PITN-GA-2010-264564. EWNG gratefully acknowledges the
support of the Wolfson Foundation, the Royal Society and the Pauli Center for Theoretical Studies. 

\bibliographystyle{JHEP}

\end{document}